\documentclass[aps,prd,twocolumn,showpacs,superscriptaddress]{revtex4}

\usepackage{amsfonts}
\usepackage{amsmath}
\usepackage[dvips]{graphicx}
\usepackage{color}

\renewcommand{\d}{d}

\begin{document}

\title{Dark matter effects in vacuum spacetime}
\date{\today}

\author{Luca Rizzi}
\email{lrizzi@dfm.uninsubria.it}
\affiliation{Department of Physics and Mathematics, Universit\`a dell'Insubria, via Valleggio 11, 22100 Como, Italy}
\author{Sergio L. Cacciatori, Vittorio Gorini}
\affiliation{Department of Physics and Mathematics, Universit\`a dell'Insubria, via Valleggio 11, 22100 Como, Italy}
\affiliation{INFN, Sezione di Milano, via Celoria 16, 20133 Milano, Italy}
\author{Alexander Kamenshchik}
\affiliation{Department of Physics, Universit\`a di Bologna, via Irnerio 46, 40126 Bologna, Italy}
\affiliation{INFN, Sezione di Bologna, via Irnerio 46, 40126 Bologna, Italy}
\affiliation{L.D. Landau Institute for Theoretical Physics, Kosygin street 2, 119334 Moscow, Russia}
\author{Oliver F. Piattella}
\affiliation{Department of Physics, Universidade Federal do Esp\'irito Santo, avenida F. Ferrari 514, 29075-910 Vit\'oria, Esp\'irito Santo, Brazil}
\affiliation{INFN, Sezione di Milano, via Celoria 16, 20133 Milano, Italy}

\begin{abstract}
We analyze a toy model describing an empty spacetime in which the motion of a test mass (and the trajectories of photons) evidence the presence of a continuous and homogeneous distribution of matter; however, since the energy-momentum tensor vanishes, no real matter or energy distribution is present at all. Thus, a hypothetical observer will conclude that he is immersed in some sort of dark matter, even though he has no chance to directly detect it. This suggests yet another possibility of explaining the elusive dark matter as a purely dynamical effect due to the curvature of spacetime.
\end{abstract}


\maketitle

\section{Introduction}

According to the so-called standard cosmological model, the dustlike matter constitutes approximately 27\% of the total energy density of the Universe (the remaining 73\% is dark energy, driving cosmic acceleration).
All astrophysical and cosmological evidence indicates that more than 80\% of the matter content of the Universe is in the form of some sort of nonluminous component (dark matter) \cite{Feng:2010gw, Giuliani}. However, after almost 80 years since the first indications appeared of the existence of such a component \cite{Oort:1932, Zwicky:1933}, the true nature of dark matter is still shrouded in deep mystery. Some fraction of the missing mass may be composed by baryonic objects which do not shine, such as neutron stars, black holes, white dwarfs, brown dwarfs, and planets. However, data from big bang nucleosynthesis and from the cosmic microwave background indicate that the overall baryon contribution, both luminous and nonluminous, cannot exceed about 20\% of the total \cite{Dar:1995gh, Alcock:2000ph, Masiero}, which leaves the remaining 80\% unexplained.

The most credited hypothesis, suggested by the open questions raised by the standard model of particle physics and the resulting need of its extension, is that the non baryonic dark matter is composed by stable, neutral, cold heavy particles which interact only very weakly with ordinary matter, such as several sorts of weakly interacting massive particles (WIMPs), axions, sterile neutrinos, and others. See \cite{Feng:2010gw} for an extensive review of the expected properties and methods of detection of such various possible dark matter candidates.

Several experiments and observations are under way in laboratories all over the world aimed at detecting signs of the existence of these elusive particles. See, e.g., \cite{Giuliani} for an updated review of the present state-of-the-art direct and indirect detection methods for dark matter. However, in spite of all these efforts, no conclusive evidence has so far turned up, though there are a few scanty signals that may be tentatively interpreted as dark matter signatures.

In support of the hypothesis of the existence of stable WIMPs as being responsible for dark matter are theoretical motivations, such as the need for a solution to the gauge hierarchy problem \cite{Feng:2010gw}, according to which these particles should have masses in the weak scale range (around $10$ GeV - $1$ TeV). Indeed, such a mass scale is just the right one for such particles to be produced with a relic density consistent with that required to explain the observed dark matter abundance and distribution (the ``WIMP miracle'').

There is also much hope that, if WIMPs are indeed there, they may soon be produced in $pp$ high energy interactions at the LHC, thus turning up as missing mass signals in the shower of particles produced in a collision.

An alternative approach to the dark matter problem is based on the modified Newtonian dynamics hypothesis (MOND). These models attempt to solve the missing mass conundrum by modifying Newton's gravitational law at large distances, without introducing new forms of matter \cite{milgrom1983modification}. One of the  relativistic realizations of the MOND paradigm is the tensor-vector-scalar gravity (TeVeS). This theory modifies gravity using a set of additional interacting fields to mimic dark matter \cite{PhysRevD.70.083509}. However, TeVeS has been recently ruled out by weak lensing observations, even though other MOND models are still  permitted \cite{Reyes:2010tr}.

Another approach which is worth mentioning is based on a modified metric theory of gravity in which the usual Einstein-Hilbert action is replaced by a fourth-order action constructed with the conformal Weyl tensor \cite{Mannheim:1992vj, Mannheim:2005bfa}. In particular, in the Newtonian limit, this theory leads to a gravitational potential in which, in addition to the $1/r$ term, there arise two more terms proportional to $r$ and to $r^2$. If the integration constants are suitably chosen, this potential leads to a satisfactory fit of the rotational curves of a large sample of galaxies without resorting to dark matter.

However, the viability of the above alternative approaches seems to be questioned by the relatively recent results of weak lensing observations of a unique cluster merger (the Bullet cluster), which point to the existence of dark matter independently of the assumptions regarding the nature of the gravitational force law \cite{Clowe:2006eq}.

Then, if we inspire ourselves with the Occam's razor principle, the failure so far to detect any of the hypothetical dark matter particles and the difficulties of the MOND-type models lead us to pose the question whether there may be any chance that the missing mass problem could be solved without introducing new particles or modifying general relativity. A possibility is that the missing mass is not actually composed of a non baryonic form of matter, but is rather a dynamical effect which occurs also in a vacuum, due to the local curvature of spacetime. Such a curvature would manifest itself as a gravitational force on test particles and light, mimicking the gravitational action of ordinary matter.

There are a few attempts pursuing this direction. For example, it has been shown in \cite{cooperstock2006galactic, Cooperstock:2006dt, Cooperstock:2007sc} that general relativistic nonlinearities can play an important role even in the context of weak gravity, indicating, in particular in idealized models of galaxy clusters that the higher-than-expected constituent velocities can be accounted for by the baryonic mass alone without resorting to additional nonluminous matter. Similarly, it has been recently proposed that at least part of the discrepancy between the luminous matter and the total mass in a typical cluster can be explained by considering the relativistic influence of the nonhomogeneous distribution of far galaxies \cite{Carati:2009us}. Another recent attempt \cite{Lusanna:2009fd} indicates the possibility that at least part of the action of dark matter could be explained as a purely relativistic inertial effect due to the non-Euclidean nature of three-space near ordinary matter. Finally, it has been proposed in \cite{Buchert:1999er,PhysRevD.79.083011,Roy:2009cj,Wiegand:2010uh,kolb2006cosmic} that dark matter (and dark energy) may simply be a manifestation of the inhomogeneous nature of the cosmological energy-momentum tensor: in these papers, the authors show that, after an averaging process, a new source emerges in cosmology that mimics both dark components, depending on the scale of interest.

In the spirit of the above attempts, we study in this letter a toy model based on a particular vacuum solution of Einstein equations in which a hypothetical ``Minkowskian'' observer, experimenting with the motion of test particles in the asymptotically flat region, concludes that he is immersed in a continuous and homogeneous (time-dependent) distribution of matter. Then, in spite of operating very sophisticated detectors in order to evidence the nature of these ``unseen particles,'' he will be frustratingly unable to register any signal: he is immersed in a vacuum and the local ``dark matter'' effect that he feels is a purely dynamical one, solely due to the curvature of spacetime.

Since our model has several bizarre features, we do not claim it provides a realistic answer to the origin of dark matter. It merely serves as an indication that dark matter effects may not necessarily be due to extra particles: in the absence of  corroborating signals one should be open to investigating alternative possibilities.

We employ geometrized units in which $c = G = 1$. The Minkowski metric tensor is $(1,-1,-1,-1)$.

\section{Model}

\begin{figure}
\centering
\includegraphics[keepaspectratio,width = .4\textwidth]{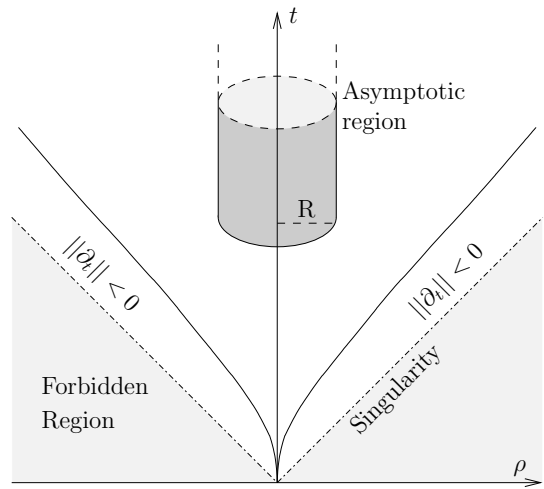}
\caption{$\rho$-$t$ section of the spacetime described by metric \eqref{metric}. The shaded, infinite cylinder represents the asymptotic region, accessible to an hypothetical ``Minkowskian'' observer. In the bands above the singularity, $\partial_t$ is spacelike.}\label{fig1}
\end{figure}

Consider the following metric
\begin{multline}\label{metric}
\d s^2 = \frac{r^3 -2m\rho^2}{r^2(r + 2m)}\d t^2 + \frac{4mt\rho}{r^2(r + 2m)}\d t \d\rho \\-\frac{r^3+2mt^2}{r^2(r +2m)}\d\rho^2 - \left(1+\frac{2m}{r}\right)\d z^2 - \rho^2 \d\phi^2\;,
\end{multline}
where $m > 0$, $t > \rho > 0$, $z \in (-\infty,+\infty)$, $\phi \in [0,2\pi)$, and $r = \sqrt{t^2-\rho^2}$. It can be obtained starting from the Schwarzschild metric
\begin{equation}
\d s^2 = \left(1-\dfrac{2m}{r}\right)\d \tau^2 - \dfrac{\d r^2}{\left(1-\dfrac{2m}{r}\right)}-r^2(\d \theta^2 +\sin^2\theta\d\phi^2)\;,
\end{equation}
and applying the complex map
\begin{equation}\label{map}
\mathfrak{J}: (\tau,r,\theta,\phi;m)\rightarrow(i\tau,ir,i\chi,\phi;-im)\;.
\end{equation}
This leads to the new metric
\begin{equation}\label{ps}
\d s^2 = \dfrac{\d r^2}{\left(\dfrac{2m}{r}+1\right)} -\left(\dfrac{2m}{r}+1\right)\d \tau^2 - r^2(\d\chi^2 +\sinh^2\chi\d\phi^2)\;,
\end{equation}
which, introducing the new coordinates
\begin{equation}
t = r \cosh\chi\;, \quad  \rho = r \sinh\chi \;,\quad z  = \tau \;,\quad  \phi  = \phi \;,
\end{equation}
takes the form \eqref{metric}. Metric \eqref{metric}, as evidenced when expressed in the form \eqref{ps} (with $m\in\mathbb{R}$), is the most general hyperbolically symmetric vacuum solution of Einstein equations. It was discovered by Harrison \cite{PhysRev.116.1285}, and some of its properties have been investigated in \cite{Gaudin:2005ek}. Note that the sign of $m$ in Eqs.~\eqref{map}-\eqref{ps} is the opposite of that considered in these references. Indeed, even though metric \eqref{ps} is a solution of the vacuum Einstein equations for any value of $m$, the sign of the latter changes some features of the metric. For example, if $m>0$ there are no horizons, whereas if $m<0$, a horizon at $r = 2 |m|$ occurs.

The nonvanishing curvature invariants of metric \eqref{metric} are
\begin{gather}
C^{\mu\nu\rho\sigma} C_{\mu\nu\rho\sigma} = \frac{48 m^2}{r^6}\;, \\
C_{\lambda\mu\nu\kappa}C^{\nu\kappa\rho\sigma} C_{\rho\sigma}^{\;\;\;\;\lambda\mu} = - \frac{96 m^3}{r^9}\;,
\end{gather}
showing it to be regular everywhere except at the singularity $r = 0$, precisely as it happens with the Schwarzschild one. However, whereas the Schwarzschild metric is static outside the horizon, metric \eqref{metric} is nonstationary in the in the region in which $\partial_t$ is timelike.

Furthermore, whereas the Schwarzschild metric is asymptotically Minkowskian in space, metric \eqref{metric} is asymptotically Minkowskian in time. Indeed, consider for the sake of definiteness an observer living in the region where $\partial_t$ is timelike, and having access only to a radially limited portion of spacetime $\rho < R$, for example (see Fig.~\ref{fig1}). Let $(t,z,\rho,\phi)$ be an event in such a region and, while keeping the coordinates $z, \rho, \phi$ fixed, let the time $t$ flow. Then, for large values of $t/2m$, \eqref{metric} becomes
\begin{equation}\label{metricas}
\d s^2 \sim \d t^2 - \d z^2 - (\d\rho^2 + \rho^2 \d\phi^2)\;.
\end{equation}
In other words, whereas a local Schwarzschild observer becomes Minkowskian if he travels sufficiently far away in space $(r\to\infty)$, an observer located at some point in space in the manifold \eqref{metric} becomes Minkowskian if he waits long enough ($t\to\infty$, hence $r\to\infty$ since $\rho$ has been fixed).

Equation.~\eqref{metricas} shows that the markers $(t,z,\rho,\phi)$ can be interpreted as cylindrical coordinates for the asymptotic or ``Minkowskian'' observer. By this interpretation, metric \eqref{metric} displays an evident cylindrical symmetry: at any fixed time $t$, the space can be imagined as an infinite cylinder, having $z$ as its axis, and which is radially limited by the singularity at $\rho_\text{max} = t$. See Fig.~\ref{fig1}.

\subsection{Particle dynamics}\label{Dyn}

Consider an observer, who lives in the Minkowskian region. He probes the presence of matter in his neighborhood by observing the motion of test masses in free fall. Their trajectories are the timelike geodesics of metric \eqref{metric}. $\partial_z$ and $\partial_\phi$ are Killing vectors, so that the quantities
\begin{equation}
l_z:= \rho^2 \dot{\phi}\;, \quad\
p_z: = \left(1 + \frac{2m}{\sqrt{t^2-\rho^2}}\right)\dot{z} \label{linearmomentumz}\;,
\end{equation}
are conserved (the dot denotes the derivative with respect to proper time). $l_z$ is the $z$ component of the angular momentum per unit mass. The second equation shows that the velocity component $\dot{z} = v_z \dot{t}$ is conserved only in the asymptotically flat region. However, planar orbits with $\dot{z}=0$ are always allowed, and these are precisely the type of motions we are interested in. Thus, a test particle in these conditions will orbit on a plane $z = \text{const}$, feeling an inward coordinate acceleration given by
\begin{multline}
\frac{\d^2 \rho}{\d t^2} = -\Gamma^\rho_{tt} -2\Gamma^\rho_{tj}v^j - \Gamma^{\rho}_{jk}v^k v^j \\ +[\Gamma^t_{tt}+2\Gamma^t_{tj}v^j+\Gamma^t_{jk}v^j v^k] v^{\rho}\;, \label{acceleration}
\end{multline}
In a ``Newtonian'' approximation, that is, for sufficiently low coordinate velocities $v^j$ \footnote{In order for this approximation to be consistent, one must consider velocities such that $v^j \ll 2m/t$, as it easily follows from an analysis of the various terms in Eq.~\eqref{acceleration}.} one has
\begin{equation}\label{Forz}
\frac{\d^2 \rho}{\d t^2} \sim - \Gamma^\rho_{tt} \sim  -\frac{m}{t^3}\rho\;,
\end{equation}
where the Christoffel symbol $\Gamma^\rho_{tt}$ is expanded at the leading order in $2m/t$. Then, a test mass in planar motion feels the same force as if it were immersed in a cylindrical and homogeneous distribution of matter, which extends throughout the whole space, with a time-dependent density
\begin{equation}\label{delta}
\delta(t) = \frac{mc^2}{G}\frac{1}{2\pi(ct)^3}\;,
\end{equation}
where now we have explicitly introduced $c$ and $G$. The mass parameter is the quantity $mc^2/ G$, whereas $m$ is the geometrical mass. The observer concludes that some form of matter is homogeneously spread in the portion of space he is living in, even though no matter is present at all. Moreover, this matter density is decreasing with the third power of time, hence he also concludes that this distribution is dynamic, as if it were linearly expanding in every direction. Similar conclusions hold from the study of the trajectories of light rays.

Of course, one might argue that similar considerations could in principle be applied to the Schwarzschild metric too, since the latter is also a vacuum solution. However, the difference between the two cases is crucial: in the Schwarzschild case the putative mass is concentrated in the origin, far away from the Minkowskian observer; in the case of metric \eqref{metric}, such ``mass'' is spread everywhere and the Minkowskian observer is immersed in it.

One can gain more physical insight about the features of the planar orbits as follows. The angular acceleration for a test-particle, under the same approximations which led to Eq.~\eqref{Forz}, is 
\begin{equation}\label{accelerationphi}
\frac{\d^2 \phi}{\d t^2} \sim -\Gamma^\phi_{tt} \sim 0\;;
\end{equation}
then, $v^{\phi} \sim $ const. To be consistent with our approximations, this constant has to be very small. Thus, the test particle rotates very slowly, with an almost vanishing angular velocity. Concerning the radial motion,
the general solution of Eq.~\eqref{Forz} is
\begin{equation}\label{Forz2sol}
\rho(t) = C_1\sqrt{\bar{t}}{\rm J}_1\left(\frac{2}{\sqrt{\bar{t}}}\right) + C_2\sqrt{\bar{t}}{\rm Y}_1\left(\frac{2}{\sqrt{\bar{t}}}\right)\;,
\end{equation}
where $\bar{t} = t /m$ and ${\rm J}_1$ and ${\rm Y}_1$ are the Bessel functions of the first and second kind, respectively. $C_1$ and $C_2$ are integration constants, which are determined by the initial position $\rho_0$ and radial velocity $v_0$. Taking the derivative of Eq.~\eqref{Forz2sol} with respect to $t$ we have, for $t\to\infty$,
\begin{equation}\label{asymptvel}
 v_\infty = v_0 - \frac{m v_0}{2 t_0} + O\left(t_0^{-3/2}\right)\;.
\end{equation}
Equation \eqref{asymptvel} represents the asymptotic radial velocity of a test particle in our approximations. Then, a hypothetical probe in such a regime experiences an attractive, though rapidly decreasing, force, which lowers its radial velocity.

\section{Discussion}

There are no indications that metric \eqref{metric} has any direct physical significance. However, it often happens that even the study of nonphysical solutions may provide insight on hidden aspects of a physical theory. In the context of general relativity, metric \eqref{metric} is a clear example. Indeed, it allows for local, dynamical effects, which do not have an analogue in classical Newtonian mechanics: a Minkowskian observer, under rather general conditions, concludes that test particles are subject to an inward force which, according to Newtonian mechanics, he is led to interpret as due to a cylindrical and homogeneous (though time-dependent) distribution of matter. However, metric \eqref{metric} is a vacuum solution and, therefore, these effects are purely ascribable to the local curvature of spacetime. As such, this ``dark matter'' is fictitious and no experiment ever will be able to detect it.

\begin{acknowledgments}
O. F. P. was supported by the CNPq (Brazil) contract No. 150143/2010-9. A. K. was partially supported by the RFBR through the grant No. 08-02-00923. 
\end{acknowledgments}

\appendix*

\section{Exact motion}

In order to obtain the effective distribution of matter of Eq.~\eqref{delta}, we have imposed rather restrictive constraints, namely (i) the motion occurs in the quasi-Minkowskian region and (ii) the test-particle travels with low velocity. Indeed, if these requirements are not fulfilled, the motion could be quite different from the one caused by a cylindrical and homogeneous distribution of matter. However, dynamical dark-matter-like effects still occur in a vacuum, though they are not described by a simple effective potential such as the one we have considered in Eq.~\eqref{Forz}.

\begin{figure}
\centering
\includegraphics[keepaspectratio,width = .4\textwidth]{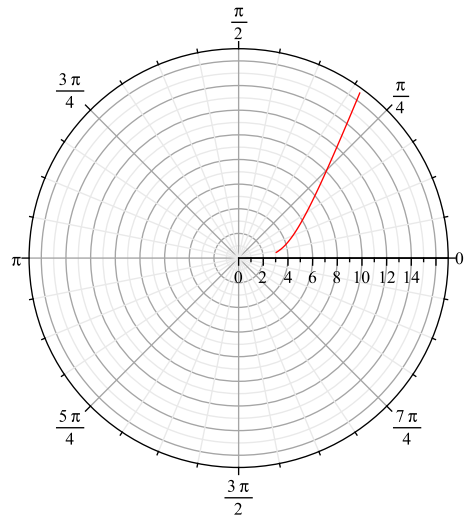}
\caption{\label{fig2} Planar timelike orbit [in $(\rho,\phi)$ coordinates]. The test-particle trajectory is bent by the spacetime curvature even though the spacetime is empty.}
\end{figure}

Thus, in this appendix, we briefly discuss the exact motion of a test particle in metric \eqref{metric}. For this purpose, it is convenient to employ coordinates $(\tau,r,\chi,\phi)$. In fact, in this case, the geodesic equations can be reduced to quadratures. For the case of planar motion (that is, $z$ constant), the trajectory of the orbit is determined by the equations
\begin{gather}
\int^\chi_{\chi_0} \frac{\d \chi}{\sqrt{\Theta(\chi)}} = \pm \int_{r_0}^r \frac{\d r}{\sqrt{R(r)}}\;, \label{a1}\\
\phi - \phi_0 = l_z \int^\chi_{\chi_0} \frac{\d \chi}{\sinh^2\chi\sqrt{\Theta(\chi)}}\;,\label{a2}
\end{gather}
where 
$R(r):=r (r+2m) (\kappa + m_0^2 r^2),$ $\Theta(\chi):= \kappa - {l_z^2}/{\sinh^2\chi}.$
$\kappa$ is an additional constant of motion, which is analogous the the total angular momentum per unit mass for the standard Schwarzschild geodesics; $m_0 = 1$ for timelike geodesics, while $m_0 = 0$ for the lightlike ones. 
A numerical example of the resulting timelike orbits, for a convenient choice of $\kappa, l_z$ and initial values of $\chi, r, \phi$, is displayed in Fig.~\ref{fig2}.

\bibliography{ps}
\bibliographystyle{apsrev}

\end{document}